\documentstyle[preprint,aps]{revtex}
\begin{document}
\preprint{SNUTP 00-012}
\draft
\title{Phase synchronization and noise-induced resonance\\ 
in systems of coupled oscillators}
\author{H. Hong and M. Y. Choi}
\address{Department of Physics, Seoul National University, 
Seoul 151-742, Korea}

\maketitle
\thispagestyle{empty}

\begin{abstract}
We study synchronization and noise-induced resonance 
phenomena in systems of globally coupled oscillators, 
each possessing finite inertia. 
The behavior of the order parameter, which measures collective
synchronization of the system, is investigated 
as the noise level and the coupling strength are varied,
and hysteretic behavior is manifested.
The power spectrum of the phase velocity is also examined
and the quality factor as well as the response function
is obtained to reveal noise-induced resonance behavior.
\end{abstract}

\bigskip
\pagebreak


\section{Introduction}

In recent years the networks of coupled nonlinear oscillators 
have attracted much attention: They serve as a prototype model 
for a variety of self-organizing systems in physics, chemistry, biology, and 
social sciences, and exhibit the remarkable phenomena of 
synchronization~\cite{Winfree,Walker69}. 
Among those the system with global coupling has been mostly studied 
both analytically and numerically,
owing to analytical simplicity and some physical as well as biological
applications~\cite{Kuramoto,Stro89,Arenas}.
In such a system of globally coupled oscillators, effects 
of nonzero inertia and of noise as well as effects of periodic driving 
on synchronization have been examined~\cite{MYChoi94,noisepaper}.
Of particular interest in the presence of noise is the possible
amplification of the response of the system, 
arising from the interplay between the noise and driving~\cite{SR:rev}.
Such stochastic resonance phenomena, having various potential applications,
have received much attention~\cite{Fauve83,McNamara89,Fronzoni93,Zhou90}.
Recently, the interesting possibility of stochastic resonance in systems 
without external periodic driving has been 
pointed out~\cite{Gang,Marchesoni,Longtin}.
For example, noise-controlled resonance behavior in 
a periodic potential with constant driving has been discussed, and 
in the limit of low damping the inertia has been shown to play 
as a surrogate of external periodic driving~\cite{Marchesoni}.
While such noise-induced resonance behavior has been investigated  
in oscillator systems with relatively few degrees of freedom, 
typically single oscillator systems,
or in systems of excitable elements~\cite{Kurrer},
the possibility of detecting the resonance 
behavior in a system of coupled (non-excitable) oscillators has not been 
properly addressed. 

The purpose of this paper is to examine whether such noise-induced resonance 
behavior can appear in a coupled oscillator system 
with many degrees of freedom.
We thus consider the system of globally coupled stochastic oscillators, 
each possessing finite inertia, subject to constant driving force,
and investigate the behavior of the order parameter, which measures the synchronization of the system, as the noise level 
and the coupling strength are varied.
To understand the interplay of noise and driving force, 
giving rise to the possibility 
of noise-induced resonance behavior in the absence of periodic driving, 
we further consider the power spectrum of the phase velocity
as the response to the driving force, and investigate 
both the zero-frequency (dc) component and the nonzero-frequency (ac) one.
The dc component of the power spectrum,
proportional to the squared time average of the phase velocity,
measures the probability that the system, kicked by noise, eventually 
escapes out of a potential-well minimum.  Namely, it describes 
the inter-well transition.
On the other hand, the ac component describes
the intra-well oscillation behavior, 
which does not induce the escape out of the potential-well minimum.
The order parameter is observed to decrease with noise, 
manifesting suppression of synchronization, and to
display hysteretic behavior with the noise level as well as 
with the coupling strength. 
Suppression of synchronization is also reflected by
growth of the dc component of the power spectrum, corresponding to
the dispersion of the mean phase velocity, with noise.
On the other hand, it is found that the generalized susceptibility,
related to the power spectrum via the fluctuation-dissipation theorem, 
increases first as the noise grows from zero,
reaches its maximum at a finite noise level, and eventually decreases as 
the noise level is increased further. 
Such noise-induced effects are also
observed in the quality factor at appropriate nonzero frequencies, 
suggesting the presence of intra-well resonance.

This paper consists of five sections: Section II introduces the 
system of coupled oscillators, each possessing finite 
inertia, subject to random noise and constant driving force.
The self-consistency equation for the order parameter, 
which measures collective synchronization in the system 
is described.
In Sec.~III, the behavior of the order parameter with the coupling strength 
and the noise level is investigated,
which manifests hysteretic behavior at low noise levels.
Section IV is devoted to the investigation of the response of the 
phase velocity to the external driving force, focusing on
the interplay between noise and driving.  
The power spectrum of the phase velocity is revealed to exhibit
noise-induced resonance in appropriate regimes.
Finally, a brief summary is given in Sec.~V. 

\setcounter{equation}{0}

\section{System of Coupled Oscillators}

We begin with the set of equations of motion governing the dynamics of 
$N$ coupled oscillators, the $i$th of which is described by 
its phase $\phi_i$ $(i=1,2,...,N)$:
\begin{equation} \label{model}
\mu\ddot{\phi_i}+ \dot{\phi_i} + {K\over N}\sum_{j=1}^N \sin(\phi_i-\phi_j)=
 \omega_i + \eta_i (t),
\end{equation}
where $\mu$ represents the magnitude of the (rotational) inertia 
relative to the damping. 
The third term on the left-hand side of Eq.~(\ref{model}) denotes the 
global coupling with strength $K/N$, whereas
the first and the second on the right-hand side represent the constant 
driving force and the random (thermal) noise, respectively.
The driving force $\omega_{i}$ is distributed over the 
whole oscillators according to the distribution $g(\omega)$, which is 
assumed to be smooth and symmetric about $\omega = 0$.
The term $\eta_i(t)$ represents independent white noise with zero mean and
correlations
$
\langle \eta_i (t) \eta_j (t') \rangle = 2T\delta_{ij}\delta(t{-}t'), 
$
where the noise level $T\,(> 0)$ plays the role of the ``effective temperature''
of the system. 
The set of equations of motion in Eq.~(\ref{model}) describes 
a superconducting wire network~\cite{Park} and may also be regarded as 
the mean-field version of an array of resistively and capacitively 
shunted junctions, 
which serves as a common model for describing the dynamics of 
superconducting arrays~\cite{Chung}.
In these cases, the constant driving force $\omega_i$
corresponds to the direct current bias.

Collective behavior of such an $N$-oscillator system is
conveniently described by the complex order parameter
\begin{equation} \label{deforder}
  \Psi \equiv {1\over N}\sum_{j=1}^N e^{i\phi_j} 
       = \Delta e^{i\theta},
\end{equation}
where non-vanishing magnitude $(\Delta \neq 0)$ indicates 
the emergence of collective synchronization and $\theta$ gives 
the average phase.
Note that the synchronized state corresponds to the superconducting
state with global phase coherence in the case of superconducting 
networks or arrays~\cite{Park}.
The order parameter defined in Eq.~(\ref{deforder}) allows us to reduce
Eq.~(\ref{model}) into a {\em single} decoupled equation
\begin{equation}  \label{single}
\mu\ddot{\phi_i}+\dot{\phi_i} + K\Delta\sin(\phi_i{-}\theta)= \omega_i 
+ \eta_{i} (t),
\end{equation}
where $\Delta$ and $\theta$ are to be determined by imposing self-consistency.
Namely, the order parameter, defined in terms of the phase via 
Eq.~(\ref{deforder}), in turn determines the behavior of the phase via
Eq.~(\ref{single}), which depends explicitly on $\Delta$ and $\theta$.
We then seek the stationary solution with $\theta$ being constant, 
which is possible due to the symmetry of the distribution of $\omega_i$
about zero.
Redefining $\phi_i{-}\theta$ as $\phi_i$ and suppressing indices, 
we write the reduced equation of motion in the form
\begin{equation} \label{single2}
\mu\ddot{\phi}+\dot{\phi} + K\Delta\sin\phi = \omega + \eta(t),
\end{equation}
which depends explicitly on the magnitude $\Delta$ of the order parameter.

In the absence of noise ($T=0$), the self-consistency 
equation for the order parameter reads
\begin{equation} \label{nodrivingorder}
\Delta = (\frac{\pi}{2}-\frac{\mu}{2})g(0) K\Delta 
               +\frac{4}{3}\mu g(0)(K\Delta)^2 
	       +\frac{\pi}{16}g''(0)(K\Delta)^3 + O(K\Delta)^4.
\end{equation}
If the distribution $g(\omega)$ is given, the collective behavior 
of the system can thus be obtained by solving Eq.~(\ref{nodrivingorder}).
In general the quadratic term of the order 
$(K\Delta)^2$ is known to
induce hysteresis in the bifurcation diagram~\cite{MYChoi94}. 
Accordingly, it has been concluded that the non-zero inertia
tends to induce hysteresis in the bifurcation diagram of the system~\cite{noisepaper}. 

The self-consistency equation for the order parameter in the 
presence of noise, particularly at such high temperatures that
$K\Delta/T\ll 1$, 
has also been considered, yielding~\cite{noisepaper}
\begin{equation} \label{e19}
\Delta=\Delta_{+} \equiv \frac{\sqrt{cK(aK-1)}}{cK^2}
\end{equation}
with the coefficients given by the integrals
\begin{eqnarray*} 
a &=& \int_{-\infty}^{\infty} d\omega \, g(\omega) \,
      \frac{T-\mu\omega^2}{2(T^2+\omega^2)}, \nonumber \\
b &=& \int_{-\infty}^{\infty} d\omega \, g(\omega) 
     \left[~\frac{T+\mu (T^2 - \omega^2)-\mu^2 \omega^2 T}{4(T^2+\omega^2 )^2}
     +\frac{\mu^3 \omega^2 +2\mu^2 T}{8(T^2+\omega^2)}\right.\nonumber\\
&&~~~~~~~~~~~~~~~~~\left.  
-\frac{6T+\mu(8T^2-\omega^2)+\mu^2 T(8T^2-\omega^2)} 
{8(T^2+\omega^2)(4T^2+\omega^2)}~\right]. 
\end{eqnarray*}
In this case collective behavior of the system has been obtained as follows:
When $K<K_c \equiv 1/a$, only the null solution ($\Delta=0$) is possible.
At $K=K_c$, on the other hand, the null solution loses its stability 
and the nontrivial solution $\Delta_+$, together with the unphysical solution 
$\Delta_- \equiv -\Delta_+$, emerges via a pitchfork bifurcation.
Subsequently, it grows in a continuous manner 
$(a^2/\sqrt{c})(K-K_c)^{1/2}$ 
as $K$ is increased beyond $K_c$~\cite{MYChoi94,noisepaper}.

\section{Phase Synchronization}

In this section we present in detail the behavior of the order parameter with 
the coupling strength and the noise level. 
We have performed extensive numerical simulations on the equations of 
motion given by Eq.~(\ref{model}) at various noise levels and coupling strengths.
The order parameter $\Delta$ has been computed from the 
definition given by Eq.~(\ref{deforder}), and its behavior depending on 
the coupling strength and the noise level has been examined.
In simulations Eq.~(\ref{model}) has been integrated with discrete time steps 
of $\delta t=0.001$, and for convenience, a semi-circle distribution 
of radius $r=0.5$ has been chosen for $g(\omega)$. 
(We have also considered other types of distribution 
such as Gaussian, only to find no qualitative change.)
In computing the order parameter, $N_t=10^5$ time steps 
have been used while the data from the first $5\times 10^4$ steps
discarded at each run. 
Both $\delta t$ and $N_t$ have been varied to confirm that the stationary
state has been achieved.
We have then computed the order parameter in the system of $N=2000$ oscillators,
each having the inertia $\mu=0.8$.

The obtained behaviors of the order parameter 
with the coupling strength and the noise level are displayed
in Figs.~1 and 2.
Figure 1(a) shows the behavior as the coupling strength $K$ is varied with the noise level $T$ kept fixed: 
Circles and squares describe the behavior of the order parameter 
as the coupling strength is increased and decreased, respectively.
At zero noise ($T=0$), 20 independent runs have been performed
with different initial configurations, over which averages are taken.
The corresponding error bars have been estimated by the standard deviations
whereas those data points without explicit error bars have errors 
smaller than the size of the symbol. 
Note the hysteresis manifested at zero noise and
weakening as the noise strength is increased from zero.
These characteristic features as $K$ is varied for given $T$ agree well
with the results of Ref.~\cite{noisepaper}.
Figure 1(b) shows that the critical coupling strength $K_c$, beyond which 
synchronization sets in, increases monotonically with the noise level $T$,
demonstrating the suppression of synchronization by noise. 
Here $K_c$ has been estimated by the value of the coupling strength 
at which the order parameter $\Delta$ first becomes nonzero
to the precision of $10^{-1}$.
The circles and squares, again corresponding to the data for increasing and decreasing $K$, respectively, have been obtained from
averages taken over 10 independent runs with different initial configurations,
and the error bars estimated by the standard deviation. 
Thus, unlike the excitable system~\cite{Kurrer}, noise-induced synchronization
does not emerge here.
Further, the hysteretic behavior, reflected by the difference in the
critical coupling strength between the two cases, 
is revealed to diminish conspicuously as the noise level is increased.

In Fig.~2(a) the behavior of the order parameter 
with the noise level $T$ for fixed coupling strength $K$ is displayed.
Here circles and squares represent the data for increasing and 
decreasing $T$, respectively, and 
the typical error bars, estimated by the standard deviation obtained from $20$ independent runs with different initial configurations,
are shown on the data points at $K=0.7$.
Figure 2(b) displays the detailed behavior with the noise strength 
for the coupling strength $K=0.7$, again manifesting
the hysteresis.  Note that the hysteresis is most conspicuous for the coupling
strength around this value, decreasing as the coupling strength is increased.
At weak coupling strengths, the system is not synchronized ($\Delta =0$),
giving no hysteresis. 
It is thus concluded that the system exhibits quite generally 
hysteretic behavior 
as either the coupling strength or the noise level is varied,
which has its origin in the non-vanishing inertia.

\section{Noise-induced Resonance}

In this section we examine the phase velocity and its power 
spectrum, and investigate the possibility of the noise-induced resonance.
The power spectrum of the phase velocity $\dot\phi_i$ is given by
\begin{equation} \label{ps}
S(f) = \frac{1}{N}\sum_{i=1}^{N} {|\tilde{v}_i (f)|}^{2}
     \equiv \langle\!\langle {|\tilde{v}_i (f)|}^2 \rangle\!\rangle ,
\end{equation}
where $\tilde{v}_i (f) \equiv \int dt \,e^{2\pi ift} \dot\phi_i $
is the Fourier component of the phase velocity at frequency $f$
and the average over different noise realizations is also to be 
taken.  Thus $\langle\!\langle \cdots \rangle\!\rangle$ 
stands for the average over the noise realizations as well as over the 
whole oscillators.
It is related to the response function of the system 
via the fluctuation-dissipation theorem~\cite{ccl}:
\begin{equation} \label{FDT}
S(f) = 2T \mbox{Re} \chi (f),
\end{equation}
where $\mbox{Re}$ denotes the real part and 
the generalized susceptibility $\chi (f)$ is defined to be
the Fourier transform of the appropriate linear-response function.
When the system is disturbed by (time-dependent) external driving, 
the resulting change in the average phase velocity takes the form
\begin{equation}
\delta \langle\!\langle \tilde{v}_{i} (f)\rangle\!\rangle 
 = \chi (f) \delta I(f),
\end{equation}
where $\delta I(f)$ is the Fourier component of the (uniform) 
external driving at frequency $f$.
In particular, the dc component of the power spectrum, describing the
dc response, reads 
\begin{equation}\label{powerspect}
S(f{=}0) = \frac{1}{N}\sum_{i=1}^{N}{|\tilde{v}_{i}(0)|}^2 
       = \frac{1}{N}\sum_{i=1}^{N}\left[\int dt \,\dot\phi_i \right]^2
       \propto \langle\!\langle {\langle\dot\phi_{i}\rangle }^2
       \rangle\!\rangle ,
\end{equation}
where $\langle \cdots\rangle$ denotes the time average. 
In the case of a superconducting wire network or array, the phase velocity 
can be identified with the voltage via the Josephson relation, and the
system is driven appropriately by a (time-dependent) external current.
Accordingly, Eq.~(\ref{FDT}) connects the generalized resistance
with the voltage power spectrum~\cite{ccl}.

To investigate the dc component of the power spectrum, 
we begin with Eq.~(\ref{single}) and consider two types of the solution,
depending on the coupling strength:
In the limit of weak coupling strength, each oscillator in the system
favors to oscillate with its own frequency and
the system is not synchronized, yielding $\Delta\approx 0$. 
The solution of Eq.~(\ref{single}) is then given by
\begin{equation} \label{poss}
\dot\phi_i = \omega_i + (v_{i0} -\omega_i ) e^{-t/\mu}
+ \frac{1}{\mu}\, e^{-t/\mu} \int_{0}^{t} dt' \,e^{t'/\mu}\eta_{i}(t'),
\end{equation}
where $v_{i0} \equiv \dot\phi_i (t=0)$ is the initial phase velocity.
Taking the time average of Eq.~(\ref{poss}) in the stationary state
($t \rightarrow \infty$), we obtain the mean phase velocity or the
frequency of the $i$th oscillator,
\begin{equation} \label{timeave}
\langle \dot\phi_i \rangle = \omega_i .
\end{equation}
We now take the average over the $N$ oscillators;
this reduces Eq.~(\ref{timeave}) to
\begin{equation} \label{Nave}
\langle\!\langle \langle\dot\phi_{i}\rangle \rangle\!\rangle 
 \equiv \frac{1}{N}\sum_{i=1}^{N}\langle\dot\phi_{i}\rangle 
 = \frac{1}{N}\sum_{i=1}^{N}\omega_i  
 = 0,
\end{equation}
where the symmetry of the distribution $g(\omega)$ about $\omega =0$
in the thermodynamic limit ($N \rightarrow \infty$) has been used.

On the other hand, the average of the square of 
the oscillator frequency,
corresponding to the dc component of the power spectrum, 
does not vanish:
\begin{equation} \label{squaretimeave}
\langle\!\langle{\langle\dot\phi_{i}\rangle}^2 \rangle\!\rangle 
= \frac{1}{N}\sum_{i=1}^{N} {\langle\dot\phi_{i}\rangle}^2 
= \langle\!\langle {\omega_i}^2 \rangle\!\rangle,  
\end{equation}
where $\langle\!\langle {\omega_i }^2 \rangle\!\rangle $ corresponds
to the variance of the distribution of $\omega_i $, i.e.,
$\langle\!\langle {\omega_i }^2 \rangle\!\rangle =\int 
d\omega \,g(\omega) \,\omega^2 $.
For example, in the simple case of the delta-function distribution 
$g(\omega)=(1/2)[\delta(\omega{-}\omega_0 )+\delta(\omega{+}\omega_0 )]$,
the variance is given by
$
 \langle\!\langle {\omega_i }^2 \rangle\!\rangle = {\omega_0 }^2 ,
$
whereas for the semi-circle distribution with radius $r$ we have
$\langle\!\langle {\omega_i }^2 \rangle\!\rangle = r^2 /2$.
Note that in this weak-coupling limit 
$\langle\!\langle{\langle\dot\phi_i \rangle }^2 \rangle\!\rangle$
as well as $\langle\!\langle \langle\dot\phi_i \rangle \rangle\!\rangle$
does not depend on the noise level $T$, indicating the absence of the
noise-induced effects.

In the limit of strong coupling strength, the
oscillators tend to oscillate in a coherent manner, displaying
synchronization $(\Delta\approx 1)$. 
Since the order parameter $\Delta$ in Eq.~(\ref{single})
depends explicitly on the noise,
decreasing with the noise level $T$, 
it is expected that unlike in the weak-coupling limit
$\langle\!\langle{\langle\dot\phi_{i}\rangle}^2 \rangle\!\rangle$ 
varies with the noise level.
When the noise level is sufficiently low ($T \approx 0$) in this strong-coupling
limit, the system is fully synchronized and described by 
the stationary solution
\begin{equation}
\phi_i = \sin^{-1}\Biggl(\frac{\omega_i }{K\Delta}\Biggr),
\end{equation}
which yields 
$\langle\!\langle\langle\dot\phi_i \rangle \rangle\!\rangle =0$ and  
$\langle\!\langle{\langle\dot\phi_i \rangle }^2 \rangle\!\rangle =0$. 
At high noise levels ($T\rightarrow \infty$), on the other hand, 
the system is not synchronized ($\Delta\approx 0$),
and we obtain 
$\langle\!\langle{\langle\dot\phi_i \rangle}^2 \rangle\!\rangle = 
\langle\!\langle{\omega_i }^2 \rangle\!\rangle$,
similarly to the case of the weak-coupling limit.
Accordingly, in the strong-coupling limit, 
$\langle\!\langle{\langle\dot\phi_i \rangle}^2 \rangle\!\rangle$ 
is expected to behave with noise as follows: At low noise levels, 
$\langle\!\langle{\langle\dot\phi_i \rangle}^2 \rangle\!\rangle$ 
increases from zero with the noise.
As the noise level is raised further, it saturates eventually toward its
asymptotic value, 
$\langle\!\langle{\omega_i }^2 \rangle\!\rangle$. 
Note that
$\langle\!\langle{\langle\dot\phi_i \rangle}^2 \rangle\!\rangle$ 
just corresponds to the dispersion or mean-square displacement 
of the oscillator frequencies over the system
since $\langle\!\langle \langle\dot\phi_i \rangle \rangle\!\rangle = 0$.
Its monotonic growth thus indicates suppression of
frequency synchronization, which accompanies that of
phase synchronization measured by the order parameter in Sec. III.
The dc susceptibility, given by
$\chi_0 \equiv \mbox{Re} \chi (f{=}0) = S(f{=}0)/2T \propto 
\langle\!\langle{\langle\dot\phi_i \rangle}^2 \rangle\!\rangle / T$, 
then grows as the noise level $T$ is increased from zero
and diminishes with $T$ at high noise levels;
in between it is expected to reach its maximum. 
Therefore in contrast with synchronization, which is suppressed by noise,
response of the phase velocity to the (uniform) external driving can be
enhanced by adding an appropriate amount of noise.

To confirm the analytical argument presented above, 
we have performed numerical simulations on
the set of equations of motion in Eq.~(\ref{model}).
For convenience, we have considered the semi-circle distribution 
for $g(\omega)$, and integrates Eq.~(\ref{model}) with discrete time steps 
of $\delta t=0.01$.
In computing the phase velocity, $N_t=10^5$ time steps have been used at
each run, with the data from the first $5\times 10^4$ steps discarded.
We have again varied both $\delta t$ and $N_t$ to verify that the stationary
state has been achieved, and performed 10 independent runs with different
initial configurations, over which averages have been taken. 
In this manner we have computed 
$\langle\!\langle{\langle\dot\phi_i \rangle}^2 \rangle\!\rangle$ 
in the system of $N$ oscillators,
for $N$ up to $4096$, and confirmed that 
there are no appreciable finite-size effects for $N\gtrsim 1000$.

Figure 3 presents the obtained behavior of the dc susceptibility 
or the dc component of the noise-divided power spectrum  
with the noise level $T$ in the system of 
$N=2000$ oscillators, each having the inertia $\mu=0.8$.
The semi-circle distribution of radius $r=0.5$ has been chosen for $g(\omega)$
and the coupling strength $K=3$ adopted.
In particular we have considered both cases of increasing
and decreasing the noise level, only to obtain the same results
within error bars. 
The behavior shown in Fig.~3 demonstrates that 
noise helps the system escape from the potential well,
enhancing the response of the phase velocity to external driving.
It is of interest to note that $T_m (\approx 1.4)$, at which the response
becomes its maximum, is almost the same as 
the critical noise strength $T_c$ below which synchronization sets in
[$T_c \approx 1.4 $ for $K=3$ as shown in Fig.~2(a)].
The height of the effective potential barrier of the system 
described by Eq.~(\ref{single2}) is given by $K\Delta$.
Since the order parameter $\Delta$ decreases with the noise level $T$,
the barrier height also becomes lower with $T$, helping the escape
from the potential well and enhancing the response to the
external driving.
Eventually, at $T_c$ the potential barrier vanishes and 
the response reaches the maximum.
It is thus concluded that noise not only hinders 
synchronization, making the critical coupling strength $K_c$ 
larger [see Fig.~1(b)], but also enhances the
response of the phase velocity to the external driving force. 
%

We now investigate the ac components of the power spectrum, i.e., the
power spectrum at nonzero frequencies,   
which gives the possibility of noise-induced intra-well resonance. 
For this purpose, we have also performed numerical simulations 
on the equations of motion, using the same 
parameter values, and compute
the power spectrum of the phase velocity through the use of
the fast Fourier transform. 
The obtained power spectrum as a function of the frequency 
$f$ is shown in Fig.~4.
At each noise level, averages have been taken over 10 independent runs with different initial configurations, to obtain the data represented by such 
symbols as filled circles, empty circles, filled squares, etc., and
the error bars have been estimated by the standard deviation.  
Note that in the absence of noise ($T=0$), no peak appears at any finite frequencies,
which is natural in the system without periodic (ac) driving.
When small noise comes into the system, however, a peak 
develops at a nonzero frequency ($f \approx 0.3$ in our simulation results)
and grows up with the noise, suggesting
the activation of intra-well oscillation by noise.
As the noise level is raised, the amplitude of such noise-induced 
intra-well oscillation is expected to grow, lowering its frequency.
Indeed the frequency at which the peak appears in Fig.~4
shifts toward lower values, demonstrating the noise-induced 
frequency shift.
It eventually approaches zero frequency; 
this describes the system kicked by noise in a potential-well minimum 
and escaping from the minimum.  Namely, 
the intra-well oscillation induced by noise turns into
the inter-well transition. 
To disclose the noise-induced effects in such intra-well motion,
we have also computed the generalized susceptibility at several frequencies
versus the noise level, where noise-induced enhancement in
the response can again be observed.
In particular, at finite frequencies, 
it is convenient to characterize such noise-induced effects 
by the appropriate quality factor
\begin{equation}
Q \equiv S_{max} (\delta f / f_{max})^{-1},
\end{equation}
where $S_{max}$ is the peak height of the power spectrum, 
$f_{max}$ is the corresponding frequency,
and $\delta f$ is the half-width of the peak.
Thus the quality factor $Q$, given by the ratio of the peak height to the relative
width, measures the degree of the coherent motion~\cite{Gang}.
We have computed $Q$ from the power spectrum obtained 
from 10 independent runs, taking the average at each noise level.
The obtained behavior of the quality factor $Q$ as a function of 
the noise level $T$ is shown in Fig.~5, which
demonstrates the presence of the intra-well resonance induced by noise.
The value $T \approx 0.7$ at which $Q$ reaches its maximum
is apparently lower than that for the inter-well 
motion in Fig.~3, indicating that intra-well resonance can be 
induced by weaker noise.

\section{Summary}

We have studied the synchronization phenomena and the noise-induced motion 
in a system of globally coupled oscillators, each possessing 
finite inertia, subject to constant driving force. 
The detailed behavior of the order parameter 
depending on the coupling strength and the noise level has been obtained 
from numerical simulations, which has revealed 
hysteresis both with the coupling and with the noise
as well as suppression of synchronization by noise.  
The hysteresis with respect to the coupling is most 
conspicuous in the absence of noise, weakening
as the noise comes into the system; that with respect to the noise
appears large at intermediate coupling strengths, 
diminishing with the coupling strength.
We have also considered the power spectrum of the phase velocity, 
as the response of the system to the (time-dependent) external driving,
and examined the possibility of the noise-induced resonance
in the system.
The dc component of the power spectrum, which corresponds to the dispersion
of the mean oscillator frequency, has been shown to grow with noise,
again manifesting suppression of synchronization.
On the other hand, the noise-divided power spectrum 
or the generalized susceptibility, 
which describes the response of the phase velocity to the external driving, 
has been found to display a peak at a finite noise level, revealing the presence of 
noise-induced enhancement in the response.
In particular, the noise-induced resonance in the intra-well motion
has been observed in the behavior of the quality factor with the noise strength.
%
It is thus concluded that noise in the system of coupled oscillators
not only suppresses phase synchronization
but also helps the system to escape from a potential-well minimum
in the response of the phase velocity, inducing the resonance. 
Such noise-induced resonance
may be manifested by a resonance peak of the voltage power 
spectrum in the case of a superconducting wire network.
Finally, we note that the major role of inertia is 
to bring about hysteresis in the response
of the system. 
The inertia is in general necessary for the system to possess (finite) 
natural frequencies, and expected to be essential to the ac resonance
at these finite frequencies. 
On the other hand, it may not be crucial in the dc resonance 
behavior of the power spectrum of the phase velocity in the
system of coupled oscillators.
Preliminary results we have obtained for the case without inertia
indeed indicate that the peak at zero frequency persists whereas that 
at finite frequency disappears, suggesting the presence of 
only the inter-well motion. 
The detailed investigation of this and other effects
are left for further study.
 
\section*{Acknowledgments}

We are grateful to G.S. Jeon for useful discussions.  MYC also
thanks D.J. Thouless for the hospitality during his stay at University
of Washington, where part of this work was accomplished.
This work was supported in part by the Ministry of Education of Korea through
the BK21 Program and by the National Science Foundation Grant DMR-9815932.

\pagebreak

\begin{figure}
\vspace*{16.5cm}
\includegraphics{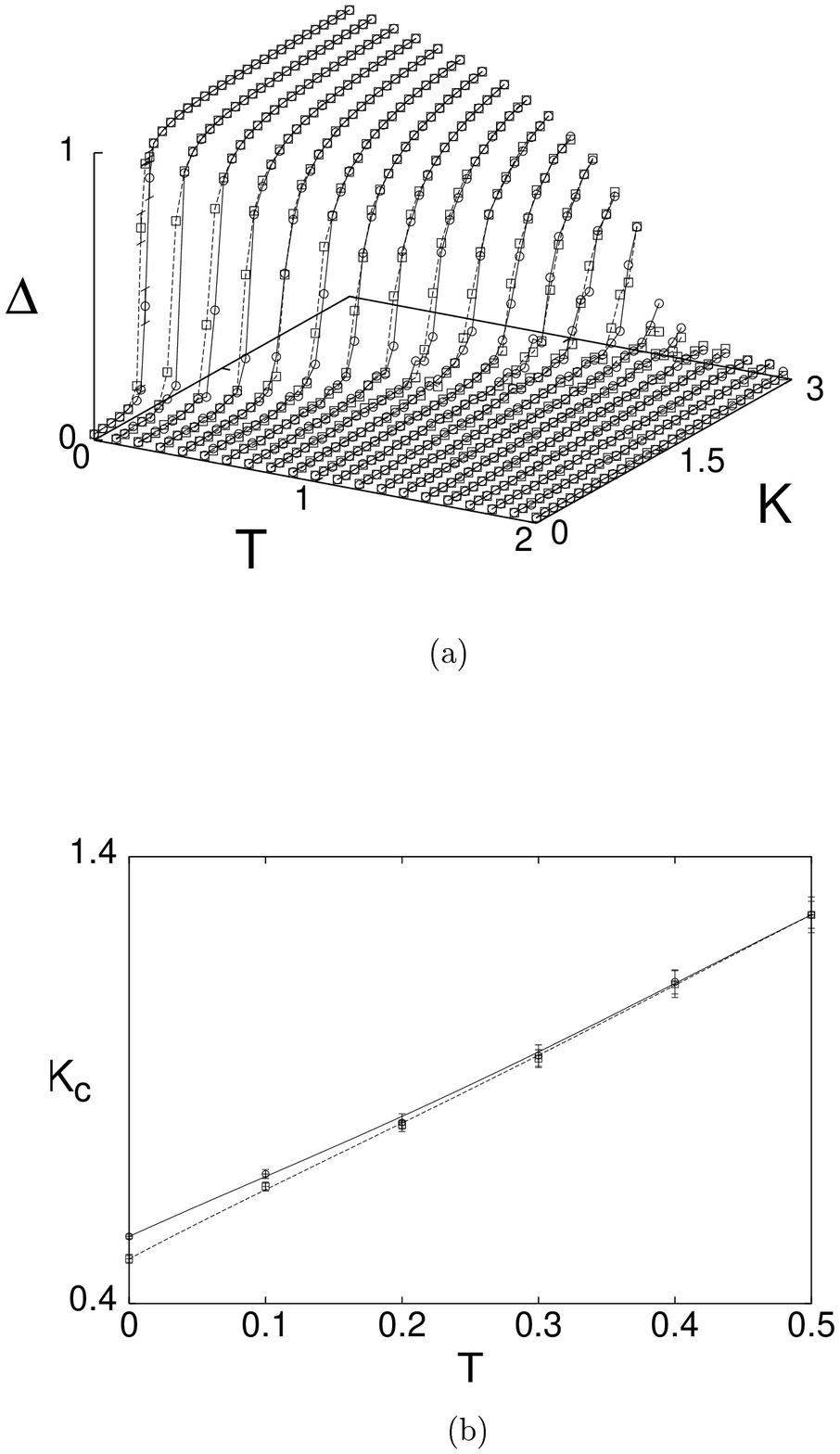}
\vspace{2cm}
\caption{(a) The order parameter as a function of the coupling strength $K$ for
various values of the noise level $T$.
Circles and squares represent the data for increasing and 
decreasing the coupling strength, respectively, and
the solid and dashed curves are merely guides to the eye.  
Hysteresis is manifested at zero noise and 
observed to weaken with the noise.
(b) Critical coupling strength, beyond which synchronization sets in, 
is shown to increase with $T$.
Notations are the same as those in (a), with the error bars
estimated by the standard deviation..
It is observed that noise in general suppresses both synchronization
and hysteresis.
}
\end{figure}

\pagebreak

\begin{figure}
\vspace*{15.5cm}
\includegraphics{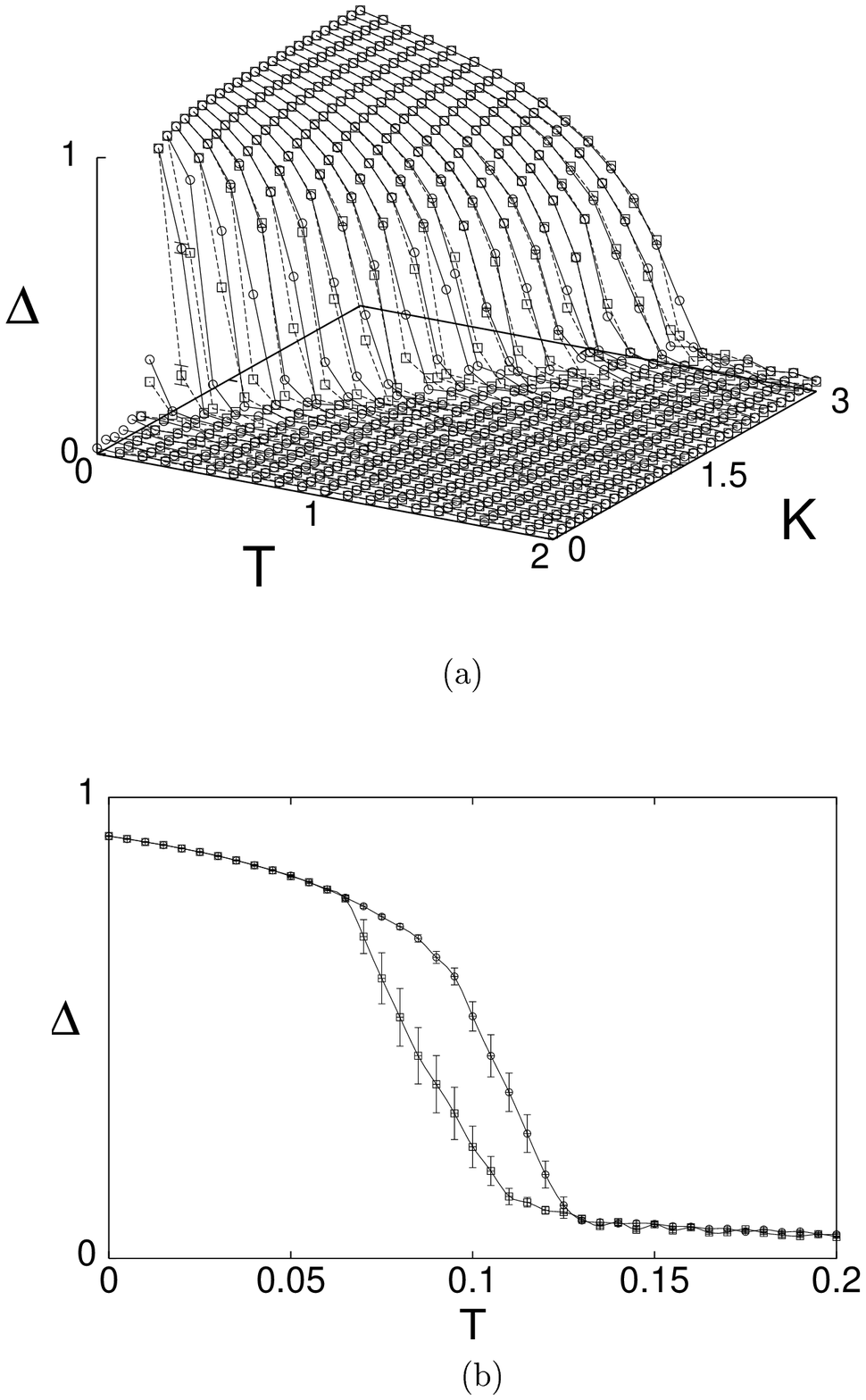}
\vspace{2cm}
\caption{(a) The order parameter as a function of the noise level $T$
for various values of the coupling strength $K$. 
Circles and squares represent the data for increasing and decreasing the
noise level, respectively, and
the solid and dashed curves are merely guides to the eye.  
(b) Behavior of the order parameter with the noise level 
at coupling strength $K=0.7$, with the same notations as in (a).
Manifested is the hysteretic behavior as the noise level is varied.
}
\end{figure}

\begin{figure}
\vspace*{15.5cm}
\includegraphics{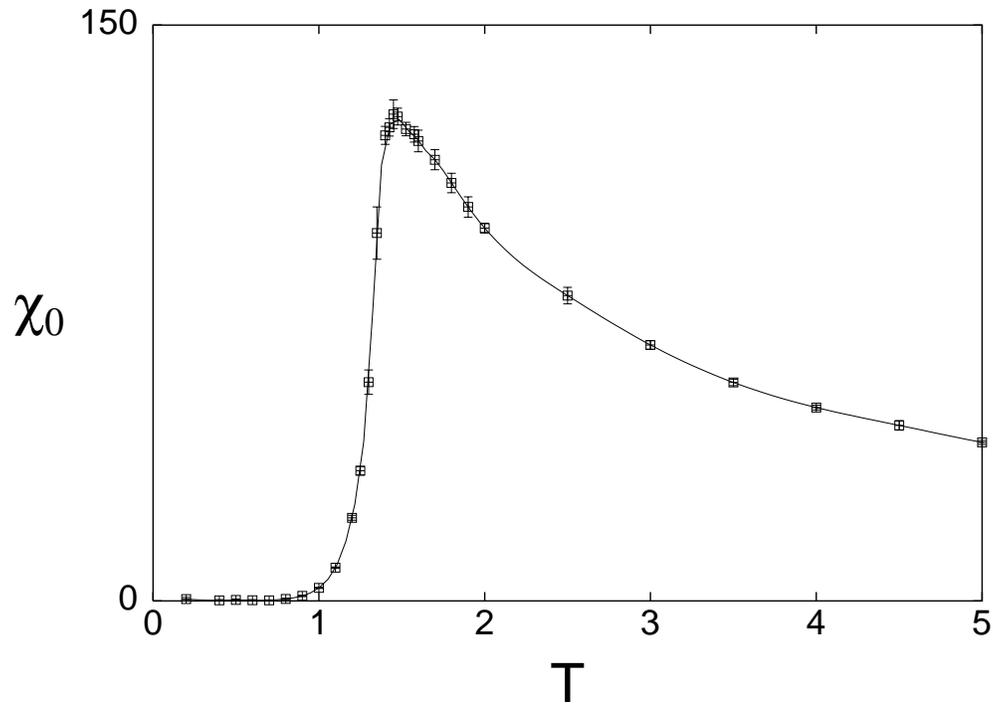}
\caption{
Behavior of the dc susceptibility $\chi_0$ (in arbitrary units)
with the noise level $T$, in the case of the
semi-circle distribution with radius $r=0.5$, revealing 
noise-enhanced response of the phase velocity. 
Error bars have been estimated by the 
standard deviation and the solid curve is merely a guide to the eye.
}
\end{figure}

\begin{figure}
\vspace*{15.5cm}
\includegraphics{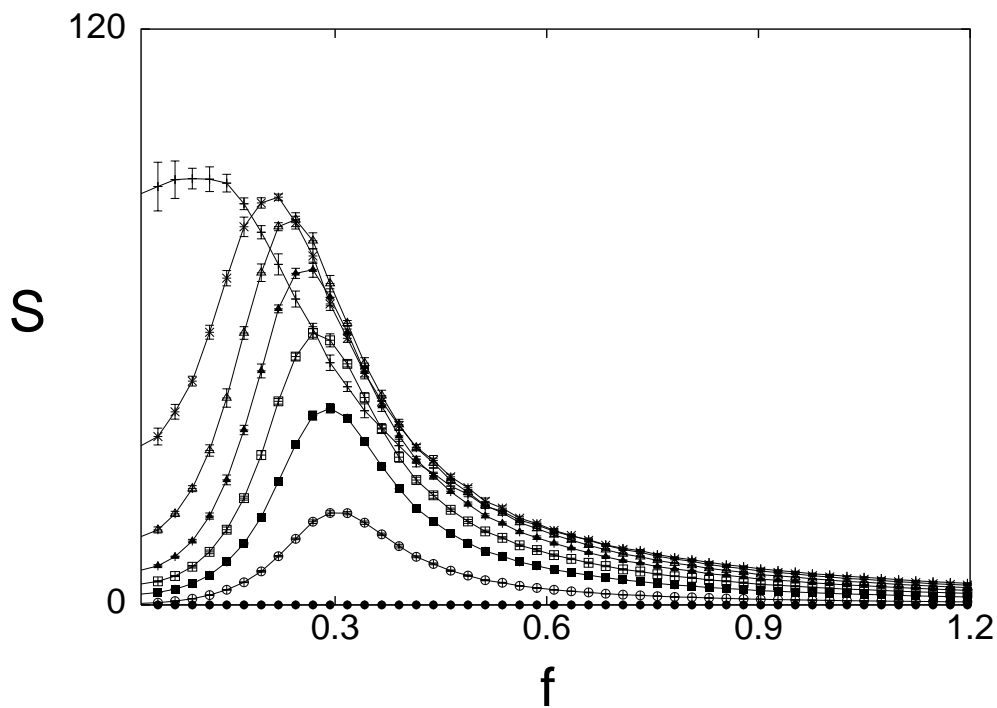}
\caption{Power spectrum of the phase velocity (in arbitrary units) at various 
noise levels:
$T=0$ (filled circles); $T=0.1$ (empty circles); $T=0.3$ (filled squares);
$T=0.5$ (empty squares); $T=0.7$ (filled triangles); $T=0.9$ (empty triangles);
$T=1.1$ (asterisks); $T=1.3$ (plus signs).
}
\end{figure}

\begin{figure}
\vspace*{15.5cm}
\includegraphics{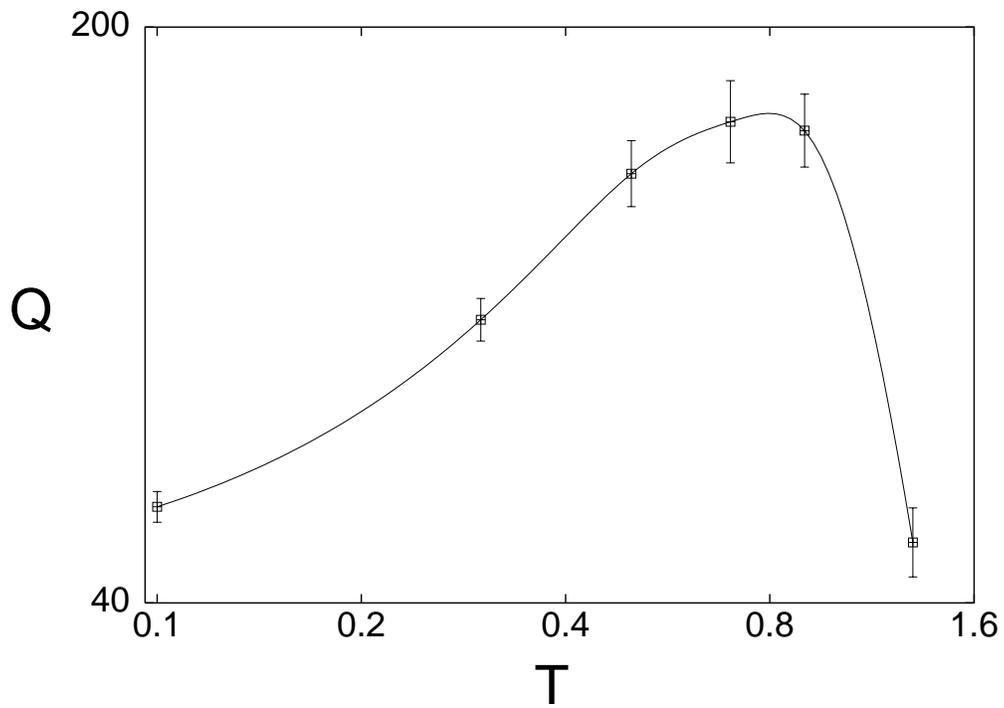}
\caption{Behavior of the quality factor $Q$ with the noise level $T$
(in the log scale), exhibiting the noise-induced resonance.
Error bars have been estimated by the standard deviation 
and the solid curve is merely a guide to the eye.
}
\end{figure}

\end{document}